\documentstyle[aps,multicol,epsfig]{revtex}

\newcommand{\be}{\begin{equation}}
\newcommand{\ee}{\end{equation}}
\newcommand{\bea}{\begin{eqnarray}}
\newcommand{\eea}{\end{eqnarray}}

\begin{document}


\title{Exact sampling from non-attractive distributions using summary 
       states}
\author{Andrew M. Childs$^1$, Ryan B. Patterson$^1$, and David J. C.
        MacKay$^2$}
\address{$^1$ Physics Department, California Institute of Technology,
              Pasadena, CA 91125, USA \\ 
         $^2$ Cavendish Laboratory, University of Cambridge, Cambridge, CB3
	      0HE, United Kingdom}
\date{8 May 2000}

\maketitle


\begin{abstract}
Propp and Wilson's method of coupling from the past allows one to
efficiently generate {\em exact} samples from attractive statistical
distributions (e.g., the ferromagnetic Ising model).  This method may
be generalized to non-attractive distributions by the use of {\em summary
states}, as first described by Huber.  Using this method, we present exact
samples from a frustrated antiferromagnetic triangular Ising model and the
antiferromagnetic $q=3$ Potts model.  We discuss the advantages and
limitations of the method of summary states for practical sampling, paying
particular attention to the slowing down of the algorithm at low
temperature.  In particular, we show that such a slowing down can occur in
the absence of a physical phase transition.
\end{abstract}


\begin{multicols}{2}
\narrowtext

\section{Introduction}

In many statistical problems, physical and otherwise, it is useful to be
able draw samples from a complex distribution.  For example, in statistical
physics one is interested in the Boltzmann distribution
\be
P(\sigma)={ e^{-\beta E(\sigma)} \over Z }
\,,
\ee
where $E(\sigma)$ describes the energy of a system in configuration
$\sigma$, $\beta$ is the inverse temperature (we set $k_B=1$), and $Z$ is a
normalizing constant (the partition function).  In general, $E(\sigma)$ may
be easy to evaluate for a particular configuration, but the number of
possible configurations makes it impractical to draw directly from the
distribution.  Yet some efficient method of sampling is desirable, as this
would allow one to calculate properties of the system that might not be
easily computed by analytical means.

In traditional Monte Carlo sampling methods, such as the Metropolis-Hastings
method~\cite{Metropolis53} and Gibbs sampling~\cite{Geman84} (also known as
the heat bath algorithm), one constructs an ergodic Markov chain whose
stationary distribution is the desired distribution.  By starting in some
state and evolving the chain for a sufficiently long time, one can
approximate a sample from the desired distribution.  Unfortunately, such a
sample is exact only in the limit of infinite time.  In practice, it is
often difficult to determine how long to wait to achieve sufficiently good
samples, and one inevitably either produces poor samples or wastes time by
running the Markov chain for longer than necessary.

However, in 1996, Propp and Wilson demonstrated the possibility of {\em
exact sampling} by the method of coupling from the past, allowing one to
produce perfect samples in a finite number of steps~\cite{Propp96}.  In the
most general case, their method requires the infeasible task of running a
Markov chain for every possible initial state of the system.  But for
certain distributions, termed {\em attractive} (such as a ferromagnetic
Ising model), Propp and Wilson showed that the task may be greatly
simplified by tracking only extremal states, permitting the practical
calculation of exact samples.  This method was generalized to
anti-attractive distributions by H\"aggstr\"om and
Nelander~\cite{Haggstrom98}.  More recently, Huber showed that one may
instead track only a single state that summarizes one's knowledge of the
system~\cite{Huber98}.  Because it does not require that the states be
partially ordered, this last method is applicable to non-attractive
distributions.

\begin{figure}
\begin{center}
\psfig{figure=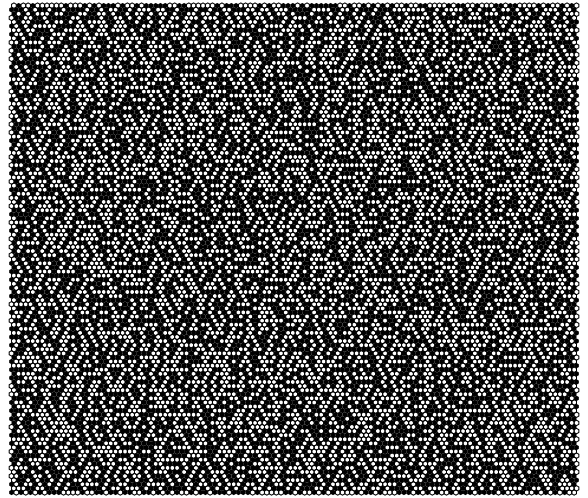,width=3in}
\end{center}
\caption{An exact sample from a triangular Ising antiferromagnet with 14,400
spins at $\beta^{-1}=4.9$ and zero applied magnetic field.}
\label{fig:isingsample}
\end{figure}

Using the summary state method, we have drawn exact samples from the
antiferromagnetic triangular Ising model and from the Potts model.
Fig.~\ref{fig:isingsample} shows one such sample.  In \S\ref{sec:methods},
we describe the methods that make this possible.  In \S\ref{sec:models}, we
briefly discuss the Ising and Potts models.  We present results from the
exact sampling of these models in \S\ref{sec:results}.  Finally, we discuss
the convergence properties of the summary state method, and we suggest
practical generalizations.

\section{Coupling from the past and the summary state method}
\label{sec:methods}

Propp and Wilson's method of coupling from the past~\cite{Propp96} is based
on the observation that, for a fixed choice of the random numbers used to
propagate a Markov chain, its possible paths in state space may ultimately
coalesce into a single trajectory.  Once two initial states lead to the same
state, they will remain in lock step thereafter.

Consider simulating a Markov chain from every possible initial state at some
fixed time $t=-T$, with the goal of taking a sample at $t=0$.  If all the
chains coalesce before $t=0$, then this finite procedure yields the same
results as a Monte Carlo simulation started at an infinite time in the past,
so the result is an exact sample.  If the chains fail to coalesce, one can
simply double the starting time to $-2T$, reusing the random numbers for the
interval $[-t,0]$ (i.e., treating the random numbers as a function of
simulation time), and repeat until coalescence is achieved.

Having to follow every possible state would make this method exponentially
intractable.  But for problems that admit a partial ordering of the states
and which are ``attractive'' --- that is, which preserve the ordering under
evolution of the Markov chain --- the computation can be vastly simplified
by tracking only the extremal states.  An example of an attractive system is
the ferromagnetic Ising model, in which it is energetically favorable for
spins to align with each other.

Huber~\cite{Huber98} and Harvey and Neal~\cite{Harvey00} have shown that the
method of Propp and Wilson may be generalized using a single {\em summary
state} instead of a pair of extremal states.  This single state summarizes
one's knowledge of the possible states of the system, allowing the state of
some subsystems to be uncertain.

For example, suppose the system is a collection of variables $\sigma_i$
taking on the values $\{\pm1\}$.  Conventional Gibbs updating sets 
\be
\sigma_i \mapsto \left\{ \matrix{
  +1 & {\rm if\ } u \le P(\sigma_i=+1|\bar\sigma_i) \cr
  -1 & {\rm if\ } u >   P(\sigma_i=-1|\bar\sigma_i) 
} \right.
\,,
\label{eq:markovupdate}
\ee
where $u$ is uniformly distributed on $[0,1]$ and $\bar\sigma_i$ denotes the
set of all variables but the $i$th.  To implement summary states, we allow
each variable to take on the additional value {\tt ?}\ which indicates
uncertainty.  We then run a modified Markov chain on this system: $\sigma_i$
is updated according to Eq.~(\ref{eq:markovupdate}) if the result is the
same for any possible assignment of $\pm 1$  to the {\tt ?}'s in
$\bar\sigma_i$; otherwise, $\sigma_i \mapsto \mbox{\tt ?}$.  As in the Propp
and Wilson method, we run the chain from successively longer times in the
past with random numbers as a function of simulation time.  When no
variables remain in {\tt ?}\ states, the algorithm has converged, and we may
take a sample at $t=0$.

For the case of attractive distributions, this procedure is exactly
equivalent to the Propp and Wilson scheme.  The value {\tt ?}\ denotes
variables that differ between the maximal and minimal states, and removal of
all {\tt ?}\ states corresponds to coalescence of the bounding chains.
However, using a single summary state, there is no requirement that the
states be ordered in any way.  Thus the summary state method can also be
applied to non-attractive distributions --- for example, the
antiferromagnetic Ising model.

Although the samples returned by this method are exact, the algorithm does
not necessarily converge after a reasonable amount of time.  Huber has shown
that for antiferromagnetic spin systems at sufficiently high temperature,
the expected running time of the algorithm is polynomial in the number of
spins~\cite{Huber98}.  However, for systems with a phase transition, the
convergence time diverges as a power law at the critical temperature, a
phenomenon known as critical slowing down~\cite{Hohenberg77}.

\section{The Ising and Potts models}
\label{sec:models}

Consider the Hamiltonian
\be
E(\sigma)= - {1 \over 2}\sum_{m,n} J_{mn} \sigma_m \sigma_n 
           - \sum_m H_m \sigma_m
\,,
\ee
where $J_{mn}$ is the coupling between spins $m$ and $n$ and $H_m$ is the
value of an external magnetic field at the location of spin $m$.  The
appropriate Markov chain update rule is Eq.~(\ref{eq:markovupdate}) with
\be
P(\sigma_i=\pm 1|\bar\sigma_i)=
{e^{-\beta E(\sigma_i=\pm 1)} \over
 e^{-\beta E(\sigma_i=+1)}+e^{-\beta E(\sigma_i=-1)}}
\,.
\ee

In the Ising model~\cite{Ising}, $J_{mn}$ is taken to be zero unless spins
$m$ and $n$ are adjacent, in which case it is some constant $J$.  Cases of
particular interest are the square lattice, in which each spin has four
neighbors, and the triangular lattice, with six neighbors per spin.  In
general, the behavior of Ising systems can vary with their spin
connectivity.  For both kinds of lattices, we use periodic boundary
conditions.  Because we may simultaneously update the states of spins whose
conditional distributions are independent, one iteration of the Markov chain
consists of two sweeps for the square lattice and three for the triangular
lattice.

In this paper, we use the normalization $J=\pm1$.  $J=+1$ corresponds to the
ferromagnetic case, in which spins prefer to point in the same direction;
$J=-1$ corresponds to the antiferromagnet.  As mentioned previously, the
ferromagnetic case is attractive.  The antiferromagnet on a square lattice
is a special case, because its properties are isomorphic to those of a
square ferromagnet.  However, for a triangular lattice, there is no such
isomorphism.  With six neighbors per spin, there is no way to minimize the
energy locally at all sites: we say the system is {\em frustrated}.

It is well known that a two-dimensional ferromagnetic Ising model exhibits a
phase transition~\cite{Onsager}.  Below a critical temperature
$\beta_c^{-1}$, there is spontaneous symmetry breaking, and the system
develops a preferred spin orientation in the absence of any magnetic field.
For a square lattice, $\beta_c^{-1}=2.27$.  At this temperature, the
relaxation time of the dynamic system diverges, a phenomenon known as
critical slowing down~\cite{Hohenberg77}.  Correspondingly, there is a
divergence in the convergence time for some Markov chain Monte Carlo
algorithms, such as coupling from the past, and exact samples cannot be
generated for lower temperatures.  Note that there is no phase transition in
the case of a triangular antiferromagnet~\cite{Wannier50}, so there cannot
be a critical slowing down in the traditional sense.

To circumvent the problem of nonconvergence below the critical temperature,
Propp and Wilson actually used a related system, the random cluster model,
to generate Ising samples~\cite{Propp96}.  Unfortunately, this model has no
obvious analog in the antiferromagnetic case.

The Potts model is a generalization of the Ising model wherein spins may
take on $q$ different values $\{0, 1, \ldots, q-1\}$~\cite{Potts52,Wu82}.
Spins interact only with others of the same type.  The Hamiltonian is
\be
E(\sigma)= -{1 \over 2}\sum_{m,n} J_{mn}
               \delta_{\sigma_m,\sigma_n}
	   -\sum_{m,k} H_m^k \delta_{\sigma_m,k}
\,.
\ee
Specifically, we consider the antiferromagnetic Potts model with $q=3$ on a
square lattice with zero magnetic field.  This model has a critical point
only at $\beta^{-1}=0$, so there is no phase transition~\cite{Ferreira94}.

\section{Results}
\label{sec:results}

\subsection{Ising model}

By implementing the summary state method, we have produced exact samples
from the Ising and Potts models.  For example, Fig.~\ref{fig:isingsample}
shows a sample from a triangular Ising antiferromagnet consisting of $120^2
= 14,400$ spins at $\beta^{-1}=4.9$ with zero applied magnetic field.

\begin{figure}
\begin{center}
\psfig{figure=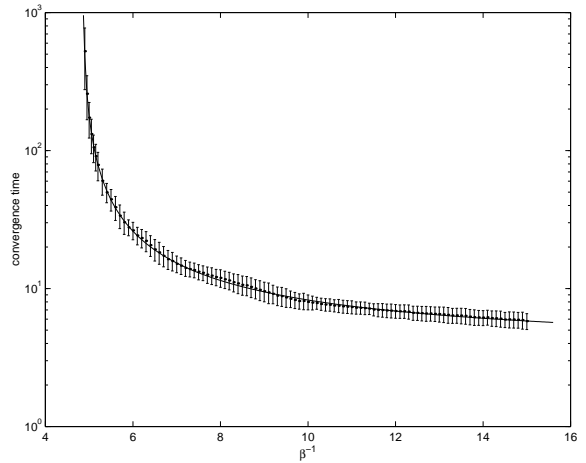,width=3in}
\end{center}
\caption{Variation with temperature of the number of iterations required for
convergence of the summary state algorithm for the triangular,
antiferromagnetic Ising model.  Each point corresponds to either 500, 1000,
or 1500 exact samples, with more samples taken at lower temperature.  The
solid line shows the fit to Eq.~(\ref{eq:fit}).} 
\label{fig:isingdiv}
\end{figure}

We find that the number of iterations required for the algorithm to converge
diverges at a threshold temperature.  We have studied this divergence using
a lattice of $N = 63^2 = 3969$ spins.  Simulations using larger $N$ (e.g.,
$N=99^2$) suggest that the outcome is not significantly affected by choosing
a larger grid size.  Fig.~\ref{fig:isingdiv} shows the divergence, to which
we have fitted a power law of the form
\be
t = {a \over (\beta^{-1} - \beta^{-1}_t)^b} + c
\,.
\label{eq:fit}
\ee
We find that the time diverges with an exponent $b=1.03 \pm 0.01$ at the
threshold temperature $\beta^{-1}_t=4.839 \pm 0.005$.

This divergence is an important feature of the summary state method.  It is
qualitatively similar to critical slowing down, but note that no physical
phase transition is involved.  In divergent situations the augmented Markov
chain has a metastable set of distributions with many {\tt ?}'s, such that
it is very unlikely for it to enter a state with no {\tt ?}'s.

To draw an exact sample using the summary state method, the system must go
from a completely uncertain state to a completely certain state.  Thus, it
must pass through a state with only a few scattered {\tt ?}'s.  For
temperatures sufficientlty near the threshold, where we know that such a
sparse configuration {\em can} be reached, we might expect that the limiting
factor is the probability that an isolated {\tt ?}\ can cause divergence.

Therefore, as a very rough estimate, we might suppose that the divergence
occurs when the probability of a single {\tt ?}\ turning one of its six
neighbors into a {\tt ?}\ rises above $1 \over 6$.  We expect that the
neighbors of any given spin $\sigma_i$ should be (on average) half up and
half down.  Replacing one of these neighbors by a {\tt ?}, we may assume the
configuration $(\uparrow\uparrow\uparrow\downarrow\downarrow\mbox{\tt ?})$
without loss of generality.  Then the threshold temperature is determined by
\be
1-{1 \over 1+e^{4\beta}}-{1 \over 2} = {1 \over 6}
\,,
\ee
which has the solution $\beta^{-1}=4/\ln 2 \approx 5.8$.

\begin{figure}
\begin{center}
\psfig{figure=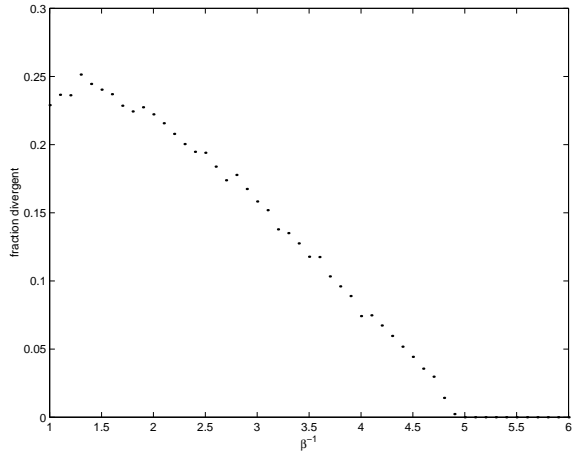,width=3in}
\end{center}
\caption{Fraction of equilibrium systems that diverged after a single {\tt
?}\ was added.  The data are from 9000 trials at each temperature.}
\label{fig:querydiv}
\end{figure}

To examine the validity of a threshold temperature analysis based on the
persistence of single {\tt ?}'s, we compiled statistics on the stability of
an equilibrium system with a single {\tt ?}\ added.  Because we cannot
create exact samples for much of the temperature range of interest, we
generated approximate samples by simulating for fixed time (100 iterations)
a random initial state.  We then set one spin to {\tt ?}\ and simulated the
system forward.  If any uncertainty remained after 500 iterations, we said
the system diverged.  Fig.~\ref{fig:querydiv} shows the fraction of
divergent trials for various temperatures.  As one would expect, this
fraction goes to zero very near the threshold temperature.

\begin{figure}
\begin{center}
\psfig{figure=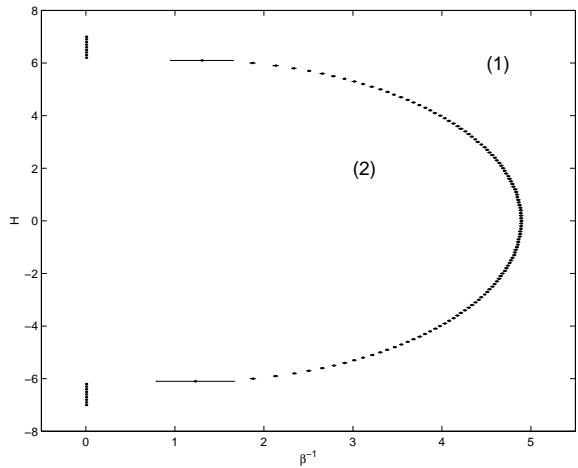,width=3in}
\end{center}
\caption{Region of convergence for the summary state algorithm on the
triangular antiferromagnetic Ising model.  Points in region (1) allow
convergence, whereas points in region (2) are inaccessible to the algorithm.
Each data point corresponds to 45 searches for the threshold, each using a
different set of random numbers.}
\label{fig:isingconv}
\end{figure}

It is also interesting to consider how the algorithm behaves when a uniform
nonzero magnetic field $H$ is applied.  Biasing the spins makes it easier for
them to choose a particular orientation, so we would expect convergence to be
easier.  Fig.~\ref{fig:isingconv} shows the region of convergence in the
$(\beta^{-1},H)$ plane.

\subsection{Potts model}

\begin{figure}
\begin{center}
\psfig{figure=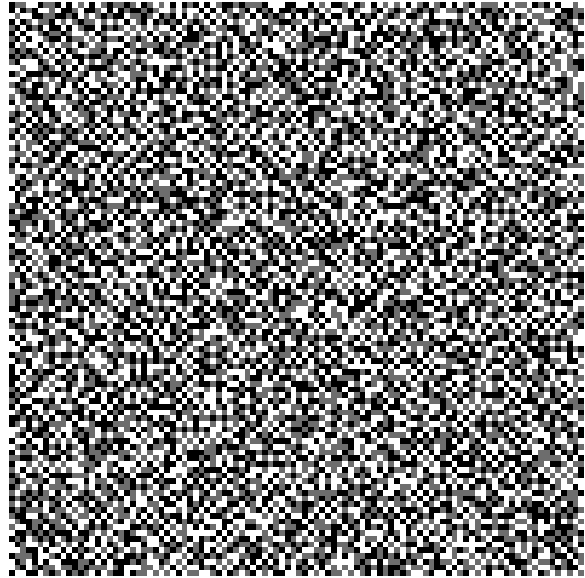,width=3in}
\end{center}
\caption{An exact sample from the $q=3$ antiferromagnetic Potts model for a
10,000 spin square lattice at $\beta^{-1}=1.2$.}
\label{fig:pottssample}
\end{figure}

In addition, we have implemented exact sampling of the Potts model for
arbitrary $q$.  Fig.~\ref{fig:pottssample} shows an exact sample with $q=3$
for a square antiferromagnetic lattice of $100^2=10,000$ spins.

A na{\"\i}ve implementation of the summary state method would augment the
possible spin values with a single {\tt ?}.  We refer to this method as
algorithm $A$.  However, it is possible to retain more information about
uncertain spins: for each spin, we store a binary $q$-bit vector $(b_1, b_2,
\ldots, b_q)$, $b_i \in \{0,1\}$.  Bit $b_i$ is set to one if it is possible
for the spin to take on the value $i$: thus the initial state of each spin
is $b=(1, 1, \ldots, 1)$.  In updating the state of the system, we set
$b_i=0$ only when the spin cannot take on the value $i$ for any allowed
configuration of its neighbors.  We refer to the latter method as algorithm
$B$.

\begin{figure}
\begin{center}
\psfig{figure=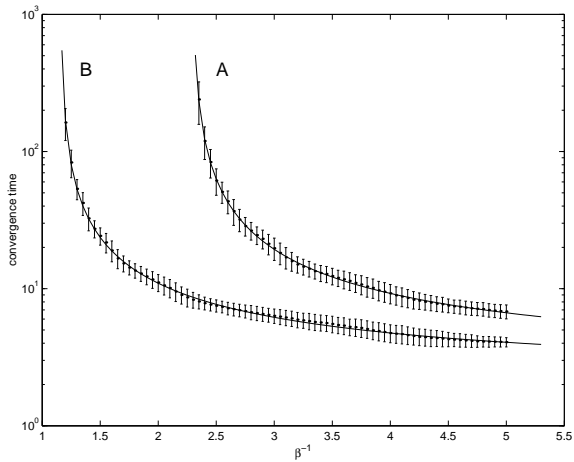,width=3in}
\end{center}
\caption{Temperature dependence of the convergence time for the square,
antiferromagnetic $q=3$ Potts model under algorithms $A$ and $B$.  Each
point corresponds to 1600 exact samples on a 4096 spin lattice.  The solid
lines give fits to Eq.~(\ref{eq:fit}).}
\label{fig:pottsdiv}
\end{figure}

To demonstrate the advantage of retaining more information in the summary
state, we have studied the convergence properties of both algorithms.  This
comparison is shown in Fig.~\ref{fig:pottsdiv}, based on data for a square
$64^2=4096$ spin lattice.  As in the Ising study, both algorithms lead to a
power law divergence with an exponent of one ($b_A=1.04 \pm 0.03$, $b_B=0.99
\pm 0.02$).  However, the threshold temperatures for the two algorithms are
quite different: $\beta^{-1}_{t,A}=2.293 \pm 0.005$, whereas
$\beta^{-1}_{t,B}=1.157 \pm 0.004$.  As in the Ising example above, neither
of the divergences corresponds to a physical phase transition.

\section{Conclusions}

We have demonstrated the usefulness of the summary state method for exact
sampling from non-attractive distributions.  In both the antiferromagnetic
Ising and Potts models, the method works above a certain threshold
temperature, with a power law divergence in the coalescence time at the
threshold.  Although similar to the phenomenon of critical slowing down,
this divergence does not occur at a physical phase transition.  As the Potts
example shows, the location of the divergence is a feature of the specific
implementation of the summary states, not of the underlying distribution.
We have shown that retaining more information in the summary state will
allow convergence at lower temperatures.

Based on this result, we may propose an improved algorithm for the
triangular antiferromagnetic Ising model.  At lower temperatures, the system
should be increasingly ordered, and tracking this order might make it easier
to gain incremental knowledge of the state of the system.  One idea is to
keep track of correlations between spins by grouping them into hexagonal
clumps of seven, which can be used to tile the triangular lattice.  Each
tile has $2^7=128$ possible states.  In analogy to the Potts method
presented earlier (in which a $q$-bit vector represents the uncertainty
about a spin), representing each tile with a 128-bit vector would allow
individually tracking the possible arrangements of those seven spins.
Within a tile, the summary state can track anticorrelation, which we expect
to arise at low temperatures.  Each edge of a tile can be easily summarized
in a 4-bit vector for comparison with its neighboring tiles.  Each tile can
then update its summary state by considering the possible states of the
neighboring edges.

Of course, summary state sampling remains a valuable tool even if it cannot
be done below some threshold.  Where it does work, it is exact.  The method
also has the advantage that we need not even know if the distribution in
question is attractive --- it can be applied to any system.

\section{Acknowledgments}
We wish to thank Radford Neal and David Wilson for several helpful
discussions.  AMC and RBP acknowledge the support of the Caltech Cambridge
Scholars Program.  DJCM's group is supported by the Gatsby Charitable
Foundation and by a Partnership Award from IBM Z\"urich.


\end{multicols}
\end{document}